\documentclass[twocolumn,citeautoscript,aps,a4paper,superscriptaddress,prl,reprint]{revtex4}
\usepackage{graphicx}
\usepackage{subfigure}
\usepackage{latexsym}   
\usepackage{amsmath,amssymb,amsthm}

\setcitestyle{super}

\usepackage{color}

\begin{document}

\title{Amyloid Fibril Solubility}

\author{\firstname{L.} G. \surname{Rizzi}} %
\affiliation{School of Chemistry, University of Leeds,  Leeds LS2 9JT, United Kingdom}
\author{\firstname{S.} \surname{Auer}} %
\email{S.Auer@leeds.ac.uk}
\affiliation{School of Chemistry, University of Leeds,  Leeds LS2 9JT, United Kingdom}

\date{\today}


\begin{abstract}
	It is well established that amyloid fibril solubility is protein specific, but how solubility depends on the interactions between the fibril building blocks
is not clear. 
	Here we use a simple protein model and perform Monte Carlo simulations to directly measure the solubility of amyloid fibrils as a function of the interaction between the fibril building blocks. 
	Our simulations confirms that the fibril solubility depends on the fibril thickness and that the relationship between the interactions and the solubility can be described by a simple analytical formula.
	The results presented in this study reveal general rules how side-chain side-chain interactions, backbone hydrogen bonding and temperature affect amyloid fibril solubility, which might prove a powerful tool to design protein fibrils with desired solubility and aggregation properties in general.
\end{abstract}



\maketitle

\section{Introduction}

	The maintenance of proteins in their soluble state is crucial to protein homeostasis in living organisms because small changes in their concentration can lead to amyloid-related diseases\cite{vendruscolo2011cshpb,ciryam2013cell}.
	At a given temperature, the solubility $C_e$ (also known as the ``critical concentration''~\cite{hillbook,powers2006biophysj,lee2009pre,gillam2013jphyscondmatt}) is the concentration of monomers $C$ at which the solution is saturated, for then the fibrils in solution cannot lengthen or shorten.
	Its importance for example in protein homeostasis arises, because the fibril solubility determines the concentration at which the fibrils can form, as well as the fibrilallion kinetics\cite{scholtz2003protsci,gillam2013jphyscondmatt} as it is a central parameter in existing nucleation theories ({\it e.g.}~refs. \cite{jarrett1993cell,lomakin1997pnas,powers2006biophysj,kashchiev2010,kashchiev2013jacs,gillam2013jphyscondmatt}).

	Experimental measurements of fibrils solubilities are intrinsically difficult, because they can be very low\cite{jarrett1994jacs,nuallain2005biochem,garai2008jcp,hellstrand2010ACSchem,baldwin2011jacs,philip2011jpepsci,yagi2013biochi} (typically 0.1-100 $\mu$M), and are beyond the detection limit of commonly used techniques for small oligomers and fibrils\cite{harper1997arevbiochem}.
	Experiments are usually made at high concentrations where fibrils form spontaneously, or in the presence of preformed fibrils, and fibril solubilities can be determined by the ratio of the on and off rates of fibril elongation\cite{harper1997arevbiochem,nuallain2005biochem}.
	These experiments show that fibril solubilities can vary strongly with point mutations of the amino acid sequence\cite{wood1995biochem,wurth2002jmolbio,williams2006jmb}, and the fibril structure~\cite{kodali2010jbiomol,qiang2013jacs}.

	Theoretically, a statistical mechanical description of the fibril solubility can be obtained in the case of linear aggregation in terms of molecular partition functions ({\it e.g.}~refs.~\cite{hillbook,ferrone1999methods,lee2009pre,schreck2013intjmolsci}).
	Also, aggregation propensities might be predicted from protein  physicochemical properties by using knowledge-based computational approaches\cite{dubay2004jmolbiol,tango2004nat,aggrescan2007bmc,tartaglia2008jmolbiol} but they are not able to distinguish between different fibril structures.
	The thermodynamic stability of amyloid fibrils has been studied by computer simulations using full-atomistic description of proteins\cite{shea2011curropn,shea2014jcpl} but they are restricted to fibrils composed of small peptide fragments and are computationally too costly to be repeated under different conditions to obtain a full phase diagram.
	Simulations using coarse-grained models to explore the self-assembly of large peptide systems have been performed, but they often determine a kinetic rather than a thermodynamic phase diagram ({\it e.g.}~refs.~\cite{hall2004biophysj,hall2006jacs}), which have been shown to be fundamentally different\cite{ricchiuto2012jphyschemB}.
	A direct calculation of the fibril solubility has been performed by Auer {\it et al.}\cite{auer2010prl,auer2011jcp} that revealed the existence of various metastable fibrillar aggregates and that the fibril solubility depends of fibril thickness.
	However, such simulations are also too costly and have only been performed for one set of interaction parameters.
	Computer simulations using a lattice model for proteins have been performed to determine the phase diagram from the peak in the heat capacity, but the interpretation of this in terms of fibril solubility is not clear\cite{bolhuis2013prl}.
	Although lattice peptide models have been employed to investigate fibril formation and growth\cite{muthukumar2009jcp,cabriolu2012jcp,bingham2013jcp,irback2013prl,irback2015jcp}, they did not provide a general understanding of how amyloid fibril solubility is determined by protein-protein interactions.

	A theoretical relationship between the interactions of the building blocks of an aggregate and its solubility can be provided by a combination of the well known van't Hoff equation and the Haas-Drenth (HD) model\cite{cabriolu2011jmolbiol,auer2015biophysj} but its applicability has never been verified by computer simulations.
	In its integrated form, the van't Hoff equation is given by
\begin{equation}
C_e = C_r e^{-\lambda} ~~,
\label{vantHoff}
\end{equation}
where $C_r$ is a reference concentration and $\lambda=L/k_BT$ is the dimensionless latent heat of peptide aggregation into fibrillar aggregates
	The corresponding latent heat $L$ can be estimated theoretically by the HD model~\cite{haas-drenthJCG1995} for protein crystals,
	and $\lambda$ is given by half the binding energy of peptides in the aggregates.

\begin{figure}[!t]
\centering
\includegraphics[width=0.42\textwidth]{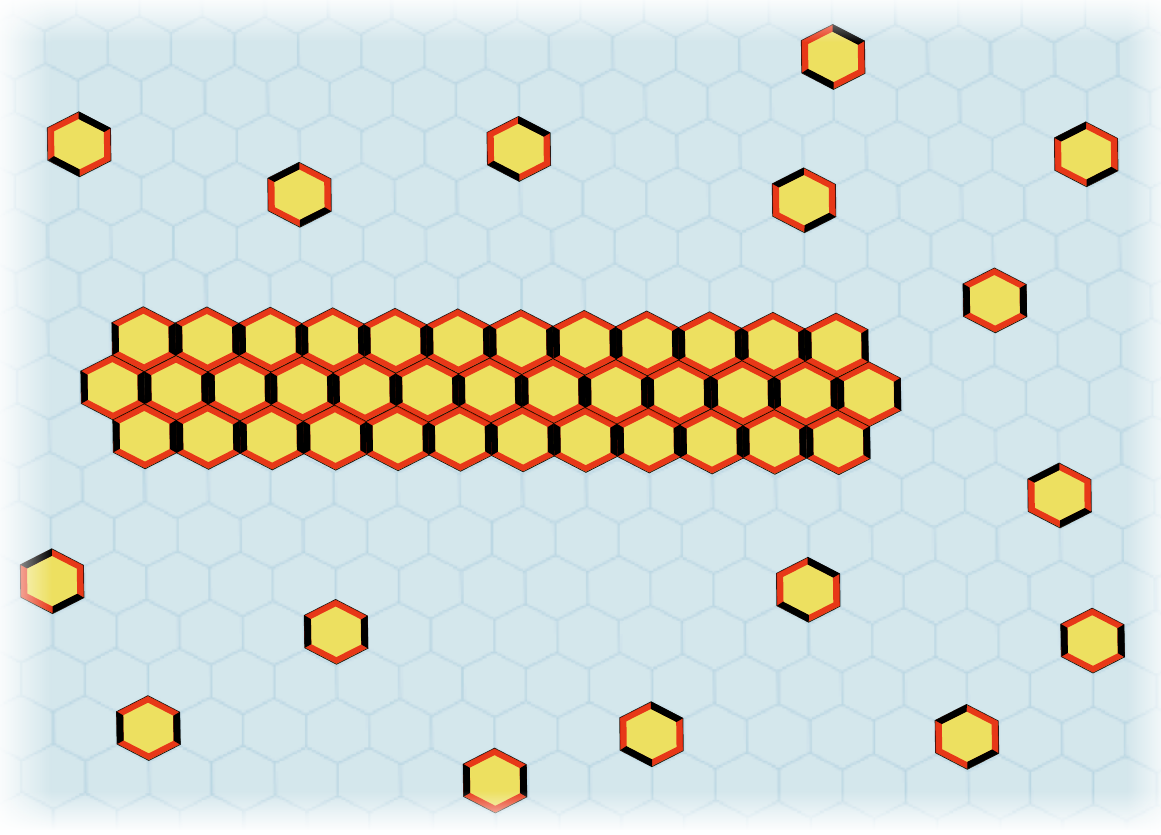}
\caption{
Illustration of an amyloid fibril composed of three $\beta$-sheets. The fibril building blocks ({\it i.e.}~the peptides)
are represented by hexagons on a triangular lattice. The four red sides of the hexagons model the hydrophobicity-mediated side-chain side-chain interactions between peptides, 
while the black sides model directional hydrogen bonds between peptides.}
\label{fig:model}
\end{figure}

\section{Peptide Model}

	To test this, we consider a peptide model that was recently used to explore the formation of amyloid fibril networks\cite{rizzi2015prl}.
	In this model, all peptides are considered in their aggregation-prone state and are represented by hexagons on a two-dimensional (2D) lattice with triangular symmetry, Figure~\ref{fig:model}.
	Each hexagon has two opposing strong bonding sides (shown in black) and four weaker bonding sides (shown in red), so that they can self-assemble into a highly ordered cross-$\beta$ structure characteristic of amyloid fibrils\cite{eisenberg2007nature,fitzpatrick2013pnas}. 
	Here the strong bonding sides correspond to hydrogen bonding which drives the formation of $\beta$-sheets and the weaker bonding sides correspond to hydrophobicity-mediated side-chain side-chain interactions that drive the fibril thickening.

	The corresponding dimensionless surface energies for the strong hydrogen and weak hydrophobicity-mediated bonds can be written as
\begin{equation}
\psi = a \sigma / k_B T = E / 2k_B T
\label{psiE}
\end{equation}
and
\begin{equation}
\psi_h = a_h \sigma_h / k_B T =  E_h / 2k_B T ~,
\label{psihEh}
\end{equation}
respectively.
	Here $\sigma$ and $\sigma_h$ are the corresponding specific surface energies, and $a$ and $a_h$ are the corresponding surface areas of the peptide faces.
	The second equality results from an approximation between the surface energies and the binding energies $E$ and $E_h$ between peptides.
	The energy $E$ corresponds to the formation of directional backbone hydrogen bonds, whereas $E_h$ is the energy due to the formation of weaker bonds modeling side-chain hydrophobic interactions between peptides.
	The fibril solubility in the combined van't Hoff and HD model is then given by 
\begin{equation}
C_e = C_r e^{-2\psi - 4\psi_h} ~~,
\label{bulksolubility}
\end{equation}
where we used that the dimensionless latent heat is given by $\lambda=2\psi+4\psi_h$.

\begin{figure}[!t]
\centering
\includegraphics[width=0.4\textwidth]{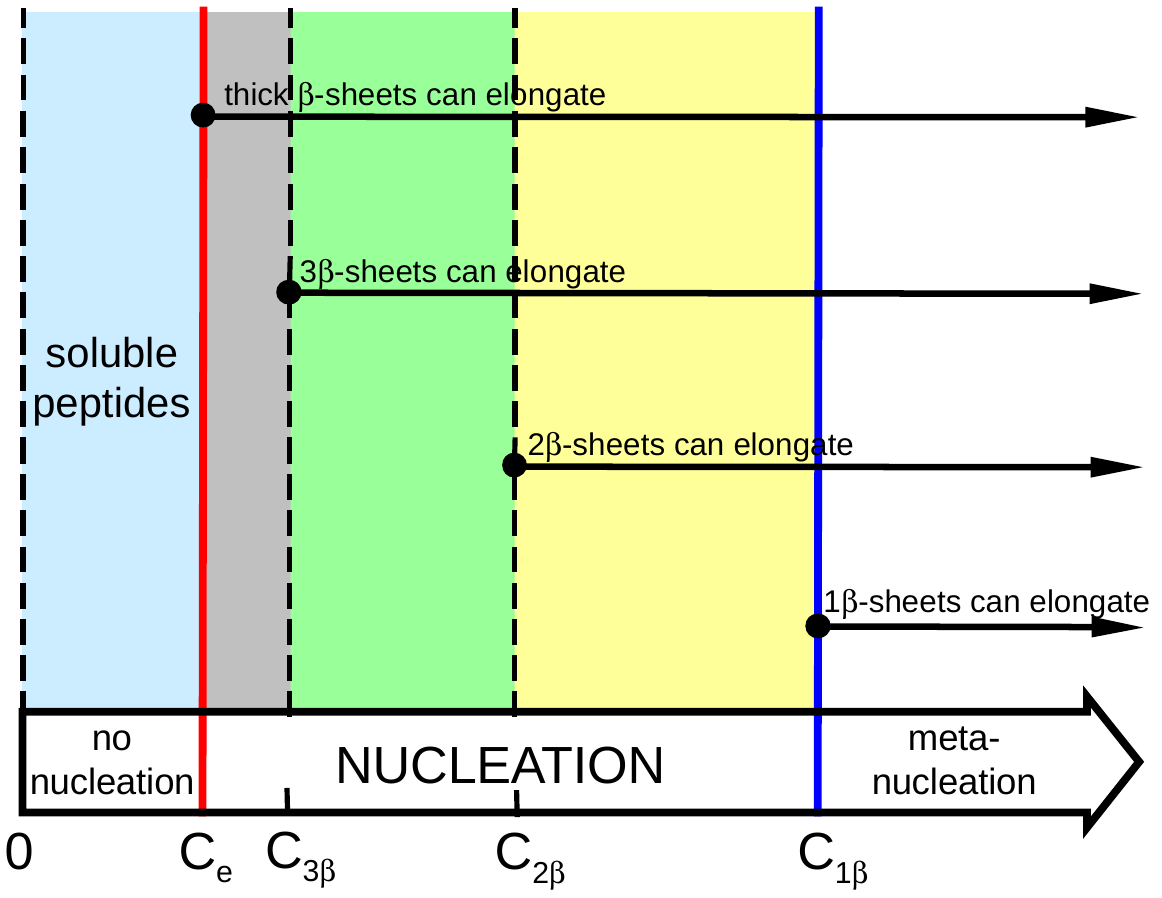}


\caption{Concentration ranges showing the fibril solubility $C_e$, the metanucleation border $C_{1\beta}$, and intermediary concentrations $C_{i\beta}$ for 
$i \geq 2$.}
\label{fig:conc_ranges}
\end{figure}

	 An important feature characteristic of amyloid fibrils is that their solubility depends on the fibril thickness~\cite{auer2010prl,auer2011jcp,cabriolu2012jcp,kashchiev2013jacs}.
	It has been shown that the solubility of fibrils composed of $i\beta$-sheet layers is given by~\cite{kashchiev2010}($i=1,2,3,\dots$)
\begin{equation}
C_{i\beta} = C_e e^{4\psi_h/i} ~.
\label{CibCePsih}
\end{equation}
Substitution of the fibril solubility $C_e$ given by Eq.~\ref{bulksolubility} into Eq.~\ref{CibCePsih} yields ($i=1,2,3,\dots$)
\begin{equation}
C_{i\beta} = C_r e^{-2\psi - 4\psi_h (1 - 1/i)} ~~.
\label{solubility:ithickness}
\end{equation}
	The analytical expressions given by Eqs.~\ref{bulksolubility} and~\ref{solubility:ithickness} define the concentration range between the fibril solubility $C_e$ and the threshold concentration $C_{1\beta}$ at the nucleation/metanucleation border (termed also ``supercritical concentration''~\cite{powers2006biophysj,gillam2013jphyscondmatt}) in which fibrils nucleate~\cite{kashchiev2010,kashchiev2013jacs}.
	As illustrated in Figure~\ref{fig:conc_ranges}, no fibrils can nucleate when the peptide concentration $C < C_e$, whereas for $C > C_{1\beta}$ the fibril nucleus is a single peptide and fibrils form in the absence of any nucleation barrier~\cite{powers2006biophysj,kashchiev2010,kashchiev2013jacs,auer2014jphyschemb,auer2015biophysj,ferrone2015jmolbiol}.
	Importantly, for intermediate concentrations, $C_e < C < C_{1\beta}$, fibril nucleation and elongation depends on their thickness~\cite{bingham2013jcp}.
	Despite the fundamental importance of those expressions, they have not yet been verified by neither simulations nor experiments.

\begin{figure}[!t]
\centering
\includegraphics[width=0.48\textwidth]{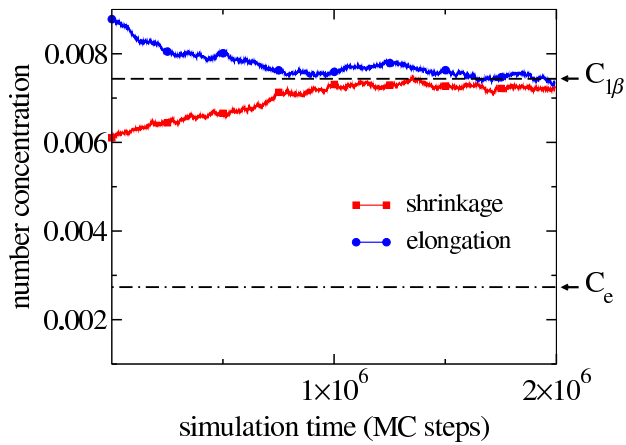}

\vspace{-0.4cm}

\caption{Solubility measurements for a 1$\beta$-sheet fibril showing that, through random attachment and detachment events, the number concentration of peptides in solution approaches the equilibrium concentration $C_{1\beta}$ instead of the solubility $C_e$.
The blue circle and red squares data were obtained from fibril elogantion and fibril schrinkage simulations, respectively.
The horizontal dashed and dash-dotted lines are theoretical values predicted, respectively, by Eqs.~\ref{solubility:ithickness} and \ref{bulksolubility} considering $\psi=3$ and $\psi_h=0.25$.}
\label{fig:sol_meas}
\end{figure}

\section{Simulation Details}

	To do this we perform Monte Carlo (MC) simulations in which we determine the equilibrium concentration at which a fibril of a fixed thickness neither grows nor shrinks.
	The simulations are performed on a 2D triangular lattice with $512^2$ sites.
	At the beginning of the simulation one single fibril of fixed thickness $i$ is put in the center of the simulation box and peptide monomers are randomly placed on the remaining lattice sites (see Figure~\ref{fig:model} for a snapshot).
	In the MC simulations we only perform diffusion-like and rotational moves as described in previous simulations~\cite{muthukumar2009jcp,rizzi2015prl}. 
	Depending on the peptide concentration the fibril will either elongate or shrink until the equilibrium concentration is reached, as illustrated in Figure~\ref{fig:sol_meas}.
	We emphasize that during the simulation the fibril can only lengthen and is not allowed to thicken.
	Typically in each simulation we perform $10^7$ MC steps and each solubility measurement is an average over four independent simulations.
	In the following we present our simulation results for the fibril solubility as a function of the hydrophobicity parameter $\psi_h$, hydrogen bonding parameter $\psi$, and temperature to verify the applicability of Eqs.~\ref{bulksolubility} and~\ref{solubility:ithickness}.

\section{Results and Discussions}

\begin{figure}[!t]
\centering
\includegraphics[width=0.48\textwidth]{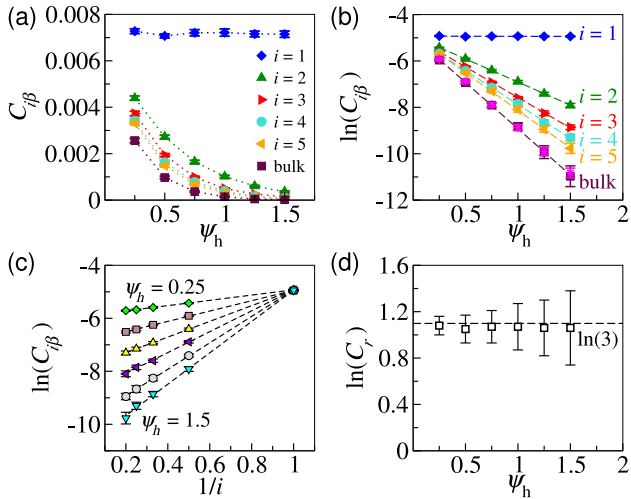}
\caption{
(a) Dependence of solubilities $C_{i\beta}$ on $\psi_h$: diamonds, up triangles, right triangles, circles, left triangles - simulation data for $1\beta$, $2\beta$, $3\beta$, $4\beta$, $5\beta$-sheet, respectively;
squares - data obtained for a bulk fibril ({\it i.e.} $i=100$); dotted lines - guide to the eyes only.
(b) monolog plot of the same data presented in (a); dashed lines - best fit of Eq.~\ref{solubility:ithickness};
asterisks - data obtained for $i \rightarrow \infty$ from scaling analysis.
(c) $\ln(C_{i\beta})$-vs-$1/i$ plot; dashed lines - best fit of Eq.~\ref{solubility:ithickness}.
(d) reference concentration $C_r$ obtained from the scaling analysis described in text; dashed line is the $C_r=3$ line. 
All results here were obtained for $\psi=3$ at room temperature, {\it i.e.} $k_B T=2.5$~kJ.mol$^{-1}$.}
\label{fig:weak_effect}
\end{figure}

	{\it Weak Hydrophobic Effect}.~The presence of hydrophobicity-mediated bonds has been shown to be an important factor in the formation of fibrillar aggregates.
	Fibril structure and point mutations change the interactions between side-chains of the peptides in the fibril and can alter their  aggregation propensity\cite{chiti2003nature}.
	In Figure~\ref{fig:weak_effect}a we present results of the $\psi_h$ dependence of $C_{i\beta}$ at fixed $\psi$. 
	As can be seen from the figure, $C_{i\beta}$ decreases with increasing $\psi_h$
for all fibrils with a thickness larger than one, and this effect becomes stronger with increasing fibril thickness.
	The values of $C_{i\beta}$ are given in units of peptides per lattice site. 
	Even though our results have been obtained in 2D, an estimate for the fibril solubilities in 3D can be obtained by scaling them as $C_{i\beta}^{3/2}$.
	Such scaling implicitly assumes that the interactions of the peptide in the third dimension, and thereby to the latent heat, are negligible.
	Assuming that the area of the lattice sites is 1 nm$^2$, the number concentrations in Figure ~\ref{fig:weak_effect}a yields threshold concentrations $C_{1\beta}$ in 3D around $1$~mM and solubilities $C_{e}$ in 3D equal to $238~\mu$M for $\psi_h=0.25$ and $0.1~\mu$M for $\psi_h=1.5$.
	Figure~\ref{fig:weak_effect}b shows the same data in a $\ln(C_{i\beta})$-vs-$\psi_h$ plot.
	A fit of these data points to Eq.~\ref{solubility:ithickness} with fixed $i$ illustrates the predicted linear dependence with slope $-4(1-1/i)$.
	Furthermore, a plot of $\ln(C_{i\beta})$-vs-$1/i$ (Figure~\ref{fig:weak_effect}c) can be used to fit Eq.~\ref{solubility:ithickness}, which provides estimates for  the solubility $C_e$ in the limit of $i\rightarrow\infty$ and the reference concentration $C_r$.
	There is good agreement between the estimates for $C_e$ obtained by this scaling analysis and the direct measurements (Figure~\ref{fig:weak_effect}b), and $C_r$ is independent of $\psi_h$ (Figure~\ref{fig:weak_effect}d). 
	We note that the value $C_r \approx 3$ (in units of peptides per lattice site) is related to the number of states that peptides can assume in our lattice model (Figure~\ref{fig:model}).
	The reason for this is that only one of the three possible orientations can lead to a successful monomer attachment to the fibril, so that there will be 3 times more monomers in solution than would be if all of them were in the same alignment of the peptides in the fibril.
	Furthermore, the fact that $C_r$ is independent of $\psi_h$ is important for theoretical studies to predict changes in the solubility as it can be assumed constant (see {\it e.g.} ref.\cite{auer2015biophysj}).
	Thus, the main effect of decreasing the hydrophobicity-mediated interactions $\psi_h$ is that it increases the solubilities $C_e$ and $C_{i\beta}$ for $i>1$, but does not alter the metanucleation border $C_{1\beta}$.
	This rationalizes the experimental observations that amino acid substitutions that decrease the hydrophobicity-mediated interaction between the fibril building blocks decrease its aggregation propensity\cite{wood1995biochem,chiti2002natstruct,wurth2002jmolbio,chiti2003nature,sanchez2006febs}.

\begin{figure}[!t]
\centering
\includegraphics[width=0.48\textwidth]{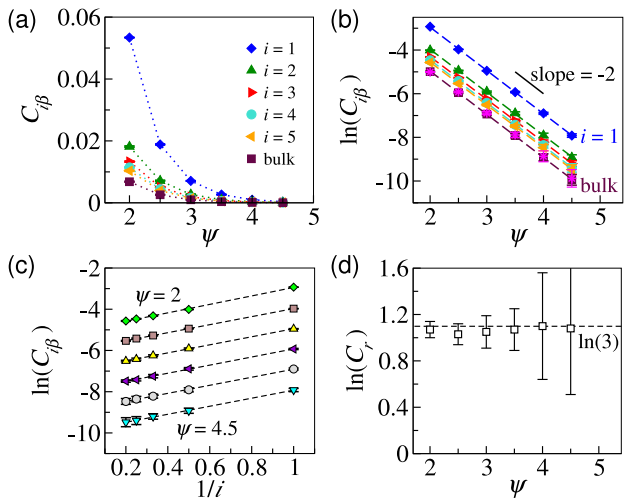}
\caption{
(a) Dependence of solubilities $C_{i\beta}$ on $\psi$. Symbols are the same as in Figure~\ref{fig:weak_effect}. Dotted lines are guides to the eyes only.
(b) monolog plot of the same data presented in (a); dashed lines - best fit of Eq.~\ref{solubility:ithickness}; 
asterisks - data obtained for $i \rightarrow \infty$ from scaling analysis as described in main text.
(c) $\ln(C_{i\beta})$-vs-$1/i$ plot for the scaling analysis; dashed lines - best fit of Eq.~\ref{solubility:ithickness}.
(d) reference concentration $C_r$ obtained from the scaling analysis; dashed line - $C_r=3$ line. All results here were obtained for $\psi_h=0.5$ and $k_B T=2.5$~kJ.mol$^{-1}$.}
\label{fig:strong_effect}
\end{figure}

{\it Hydrogen Bonding Effect}.~The lengthening of amyloid fibrils is driven by the formation of backbone hydrogen bonds in the direction of the fibril lengthening axis. 
	The strength of the hydrogen bonding network between neigboring peptides in the fibril varies depending on the amino acid sequence, because it determines the number of hydrogen bonds that can be formed~\cite{cheng2013jacs}. 
	To determine the effect of changes in the hydrogen bonding interactions we vary $\psi$ while keeping the strength of the hydrophobicity-mediated interactions $\psi_h$ fixed.
	Figure~\ref{fig:strong_effect}a shows that $C_{i\beta}$ decreases with increasing $\psi$. 
A plot of these data as $\ln(C_{i\beta})$-vs-$\psi$ coordinates, illustrates a linear dependence and that the slope is independent of the fibril thickness. Indeed, a fit of Eq.~\ref{solubility:ithickness} with fixed $i$ to these data points shows that the slope is equal to $-2$ for all $i$ (Figure~\ref{fig:strong_effect}b).
	As before, a scaling analysis to obtain the bulk solubility $C_e$ in the limit of $i\rightarrow\infty$ and the reference concentration $C_r$ can be obtained by a fit 
of Eq.~\ref{solubility:ithickness} to our data presented in a $\ln(C_{i\beta})$-vs-$1/i$ plot (Figure~\ref{fig:strong_effect}c).
	As can be seen in Figure~\ref{fig:strong_effect}b, there is again good agreement of the results obtained from our scaling analysis and direct measurements of $C_e$.
Furthermore, also in this case the so obtained reference concentration $C_r \approx 3$ is independent of $\psi$ (Figure~\ref{fig:strong_effect}d).
	The main effect of decreasing the hydrogen bond energy interaction $\psi$ is that it increases the solubility $C_e$ and the threshold concentration $C_{1\beta}$ while keeping the ratio $C_{1\beta}/C_e$ constant.
	This effect might explain experiments using hydrogen-deuterium exchange mass spectrometry to demonstrate that the solubilities of different A$\beta_{1-40}$ polymorphs increases as hydrogen bond formation between proteins decreases\cite{kodali2010jbiomol}.
\begin{figure}[!ht]
\centering
\includegraphics[width=0.48\textwidth]{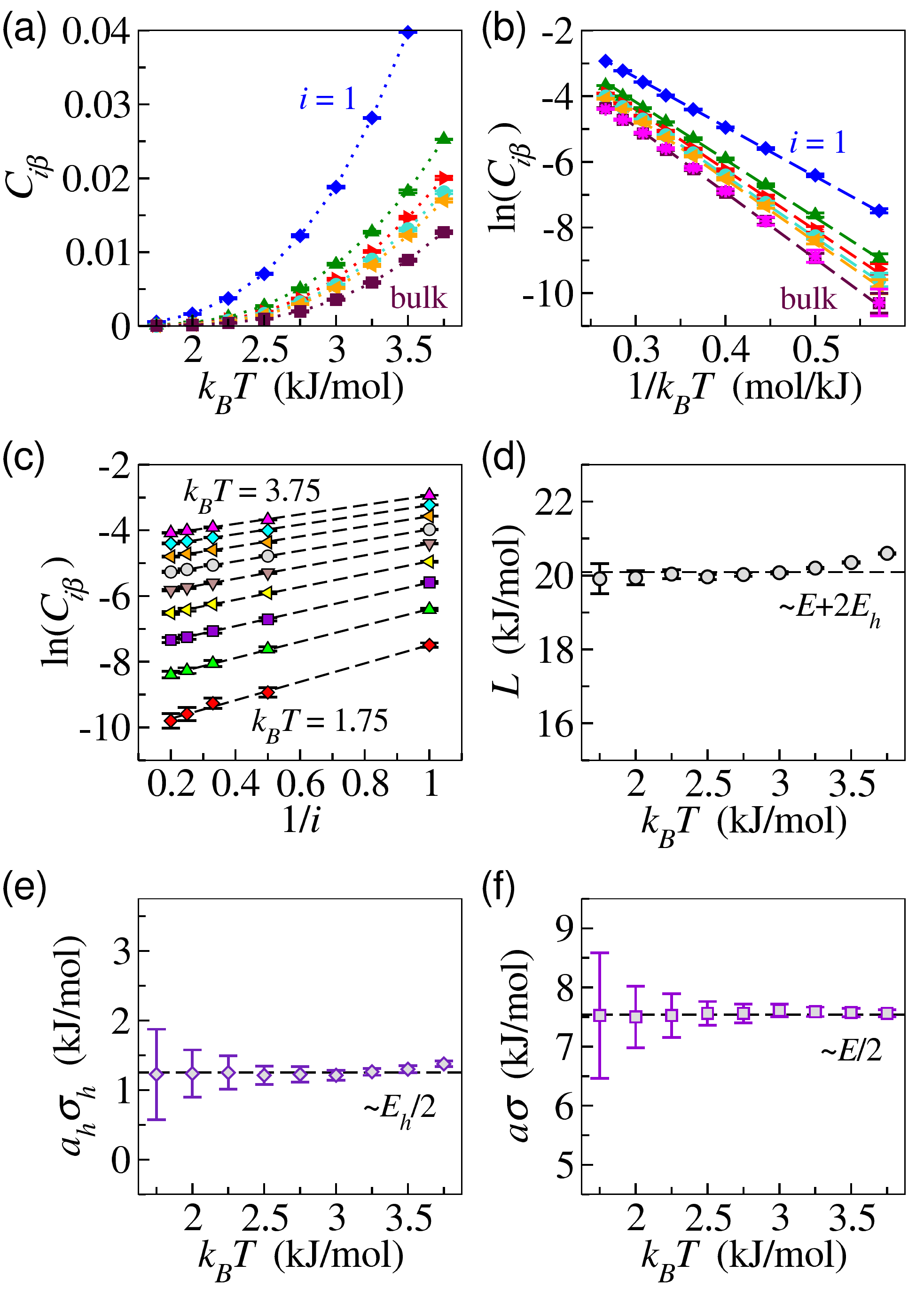}
\caption{
(a) Dependence of solubilities $C_{i\beta}$ on $k_BT$. Symbols are the same as in Figure~\ref{fig:weak_effect}. Dotted lines are guides to the eyes only.
(b) monolog plot of the same data presented in (a); dashed lines - best fit of Eq.~\ref{vantHoff} assuming $C_r=3$;
asterisks - data obtained for $i \rightarrow \infty$ from the scaling analysis described in main text.
(c) $\ln(C_{i\beta})$-vs-$1/i$ plot for the scaling analysis; dashed lines - best fit of Eq.~\ref{CibCePsih} with $C_r=3$.
(d) Specific heat $L$ extracted at each temperature from data in (c) by considering Eqs.~\ref{vantHoff} and~\ref{CibCePsih} assuming $C_r=3$; dashed line - estimate of $L$ from the slope of $\ln(C_e)$-vs-$1/k_B T$ in (b) considering Eq.~\ref{vantHoff}.
Panels (e) and (f) show, respectively, estimates for the surface energies $a_h\sigma_h$ and $a\sigma$ evaluated from the scaling presented in (a) by considering Eq.~\ref{vantHoff} and Eq.~\ref{CibCePsih} at each temperature assuming $C_r=3$; dashed lines in (e) and (f) are estimates extracted from the scaling in (b) for comparison (see text). 
All results here were obtained for binding energies $E=15$~kJ.mol$^{-1}$ and $E_h=2.5$~kJ.mol$^{-1}$.}
\label{fig:temp_effect}
\end{figure}

{\it Temperature Dependence}.
	Amyloid fibrils are known to have high thermodynamic stability, and the temperatures required for the dissociation of the fibril structure are far greater than needed for {\it e.g.}~the denaturation of native proteins~\cite{meersman2006biochm}.
	To test the effect of temperature on the fibril solubility we performed simulations in which the we fixed the binding energies and varied the temperature.
	Figure~\ref{fig:temp_effect}a illustrates that $C_{i\beta}$ increases with increasing $k_BT$, and that this effect is less pronounced with the increasing number of fibril layers $i$.
	In Figure~\ref{fig:temp_effect}b we present the same data but in a $\ln(C_{i\beta})$-vs-$1/k_BT$ plot.
	A fit of Eq.~\ref{solubility:ithickness} with fixed $i$ to these data points using a reference concentration $C_r=3$ shows a linear dependence with a slope equal to $-(E+2E_h(1-1/i))$ as predicted by Eq.~\ref{solubility:ithickness}.
	Furthermore, presenting the same data in a $\ln(C_{i\beta})$-vs-$1/i$ plot (Figure~\ref{fig:temp_effect}c), a fit of Eq.~\ref{CibCePsih} to these data points also provides estimates for the solubility $C_e$ from the intercept in the limit of $i\rightarrow\infty$.
	As shown in Figure~\ref{fig:temp_effect}b, such estimates are in good agreement with the direct measurements of $C_e$ for very thick fibrils ({\it i.e.} for $i=100$). 
In the limiting case when $i\rightarrow\infty$, the slope obtained for the bulk solubility line $C_e$ corresponds to the latent heat $L=\lambda k_BT$ (Eq.~\ref{vantHoff}), which remarkably is equal to half of the binding energies $E+2E_h$, as predicted by the HD lattice model (see Figure~\ref{fig:temp_effect}d).
	The fact that $L$ is half the binding energy cross validates our assumption that $C_r = 3$ in the linear fit of Eq.~\ref{CibCePsih} to the data presented in Figure~\ref{fig:temp_effect}c.
	Finally, the slopes obtained from the fit of Eq.~\ref{CibCePsih} to data in Figure~\ref{fig:temp_effect}c, in combination with Eq.~\ref{psihEh}, provide values for the surface energy $a_h\sigma_h$ at each temperature. As shown in Figure~\ref{fig:temp_effect}e, such values are equal to half the values of the weak bond energy $E_h/2$.
	Furthermore, combined with the estimate for the latent heat $\lambda$, the values of $a_h\sigma_h$ can be used to estimate the strong bond surface energies $a\sigma$.
	Figure~\ref{fig:temp_effect}f confirms that the values of $a\sigma$ are independent of $k_BT$ and equal to half the binding energy of the strong bond $E/2$.
	Note that the dashed lines in Figs.~\ref{fig:temp_effect}e and~f are estimates obtained from the fit of Eq.~\ref{solubility:ithickness} to data presented in Figure~\ref{fig:temp_effect} considering $i=1$. 
	The main effect of increasing the temperature is that it increases both $C_{1\beta}$ and $C_e$, which is in agreement with experimental measurements of solubilities for a short A$\beta$-model peptide\cite{hamley2010jphyschem}.
	In the example presented in Figure~\ref{fig:temp_effect}, increasing the temperature from 10\,$^{\text{o}}$C to 60\,$^{\text{o}}$C changes the solubilities $C_e$ in 3D from $25~\mu$M to $171~\mu$M, and the threshold concentration $C_{1\beta}$ in 3D from $0.61$~mM to $2.56$~mM, respectively.

\section{Conclusions}

	In summary, the main results of the analysis presented is that we numerically verified Eqs.~\ref{bulksolubility} and~\ref{solubility:ithickness} which provide simple relationships between the solubilities of fibrils of various thickness and the dimensionless specific surface energies that are given by half the dimensionless binding energies between the peptides in the fibril (Eqs.~\ref{psiE} and~\ref{psihEh}). 
	This is important, as these equations can be used to {\it e.g.} estimate changes in the fibril solubility upon point mutations or fibril structure. 
	In combination with existing nucleation theories they can also be used to predict how such changes affect their fibrillation kinetics ({\it e.g.}~refs.\cite{cabriolu2011jmolbiol,auer2015biophysj}).
	To design peptides that assemble into fibrils with desired solubility our analysis reveals some general rules:
(i) decreasing the hydrophobicity-mediated interaction $\psi_h$ will increase the solubility $C_e$, corroborating what was previously suggested by experimental studies~\cite{wood1995biochem,wurth2002jmolbio,sanchez2006febs}; 
(ii) changes in $\psi_h$ do not alter the value of $C_{1\beta}$, but the solubilities $C_e$ and $C_{i\beta}$ for $i>1$;
(iii) decreasing the hydrogen bond energy interaction $\psi$ will increase both the solubility line $C_e$ and the metanucleation border $C_{1\beta}$ while keeping ratio $C_{1\beta}/C_e$ constant;
(iv) higher temperatures increases both $C_{1\beta}$ and $C_e$.

	Finally, we note that our model is limited to peptides in an aggregate-prone state immersed in an implicity athermal solvent; thus neither conformational changes nor solvent effects\cite{hamley2010jphyschem} are considered.




L. G. Rizzi acknowledge support by the Brazilian agency CNPq (Grant N\textsuperscript{o} 245412/2012-3).



\providecommand{\latin}[1]{#1}
\providecommand*\mcitethebibliography{\thebibliography}
\csname @ifundefined\endcsname{endmcitethebibliography}
  {\let\endmcitethebibliography\endthebibliography}{}

\end{document}